\documentclass[12pt]{article}
\usepackage{amssymb,amsmath}
\begin{document}
\newcommand{\ket}[1]{\mbox{$\left| #1 \right\rangle$}}
\newcommand{\N}{\mbox{$\mathbb N$}}
\newcommand{\Z}{\mbox{$\mathbb Z$}}
\newcommand{\R}{\mbox{$\mathbb R$}}
\newcommand{\w}{\mbox{$\omega$}}
\newcommand{\QFT}{\mathrm{QFT}}

\title{The $f$-conditioned Phase Transform}
\author{\normalsize
 {\bf Dong Pyo Chi}\thanks{E-mail address: \texttt{dpchi@math.snu.ac.kr}, Fax: +822-887-4694},
 \ {\bf Jinsoo Kim}\thanks{E-mail address: \texttt{jskim@math.snu.ac.kr}},
 \ {\bf and}
 {\bf Soojoon Lee}\thanks{E-mail address: \texttt{level@math.snu.ac.kr}}\\[5mm]
\normalsize Department of Mathematics, Seoul National University\\
\normalsize Seoul 151-742, Korea
}
\date{February 25, 2000}
\maketitle

\begin{abstract}
We present a quantum algorithm for the $f$-conditioned phase transform
which does not require any initialization of ancillary register.
We also develop a quantum algorithm that can solve
the generalized Deutsch-Jozsa problem by a single evaluation of a function.
\end{abstract}

\section{Introduction}
Several quantum algorithms have been implemented by NMR quantum
computers
\cite{Chuang1,Chuang2,Collins1,Collins2,Dorai1,Dorai2,JM1,JM2,JM3,Linden,Marx,Weinstein,Yannoni}
among which much attention has been paid to the Deutsch-Jozsa
algorithm \cite{DJ} due to its simplicity whereas the power of a
quantum computer over a classical one can be demonstrated.
In NMR implementation for Deutsch-Jozsa algorithm
there are two approaches,
one of which is the Cleve's version \cite{Cleve1}
that requires an $n$-qubit control register for storing function
arguments and a one-qubit ancillary register for function evaluation
to solve the $n$-bit Deutsch-Jozsa problem.
It has been implemented by several research groups
\cite{Chuang1,Linden,Marx} up to four qubits
following the first successful implementation \cite{JM1} of a
quantum algorithm on any physical system with two qubits.
In general for a quantum computer with a larger numbers of qubits
the associated requirement of appreciable coupling between
any pair of spins raises difficulties.
The conditional phase transform enables us to eliminate
some ancillary register in the description of quantum algorithms,
one of which examples is
the refined Deutsch-Jozsa algorithm in \cite{Collins1}.
Its realization has been reported
with three-qubit arguments \cite{Collins2,Dorai2}
and with four-qubit arguments \cite{Marx}.

To perform conditional phase transform
we have to evaluate a given function on a quantum computer.
Unitary evolution property of quantum computation
necessitates at least one ancillary register
from which
we have to extract relative phases conditioned on a function.
If we can initialize the ancillary register
the phase-encoded information can easily be accomplished.
We generalize the conditional phase transform
to have arbitrary relative phases controlled by a given function,
which we call the $f$-conditioned phase transform.
We present an algorithm to implement
the $f$-conditioned phase transform
without initializing the ancillary register.
Furthermore, the application of the algorithm
turns the ancillary register back to its initial state.
This implies that we are free to compose this temporary register
while it is being used in another computational process
without corrupting its computation.
Our algorithm is optimal
in that it involves two $f$-dependent operations.
Because to realize the $f$-conditioned phase transform
at least two operations dependent of $f$ are necessary.
This is because we do not require any initialization
of the ancillary register.
If some kind of initialization is involved,
only one $f$-dependent operation is sufficient.
Using the $f$-conditioned phase transform
we develop a quantum algorithm that can solve the generalized
Deutsch-Jozsa problem by a single evaluation of a given function.

\section{The $f$-conditioned Phase Transform}
For $N \in \N$ we denote by $\Z_N=\{ 0, 1, \dots, N-1 \}$ the
additive cyclic group of order $N$. Let $\{ \ket{a} \}_{a\in
\Z_N}$ be the standard basis of the Hilbert space ${\cal H}_N$
representing the state of an $n$-qubit quantum register.

Given a function $f:\Z_N \rightarrow \Z_M$ where $N, M \in \N$,
the operation
$R_{k,f}:\ket{x} \mapsto \w_M^{kf(x)} \ket{x}$ plays an important
role in quantum algorithms for an appropriately chosen $k \in
\Z_M$ according to the problems and $\w_M=\exp(2\pi i /M)$ is a
primitive $M$-th root of unity. The resulting interference pattern
is used to determine global property of the function and most known
quantum algorithms rely on this {\em $f$-conditioned phase transform}.
In order for the values of a function to be encoded in
the phases we need a quantum circuit to evaluate a function.

On a quantum computer the evaluation of a function is performed by
a unitary operation $U_f: \ket{x}\ket{y} \mapsto \ket{x}\ket{y+f(x)}$.
The first $n$-qubit register we call the {\em control register}
contains the states we wish to interfere.
The second $m$-qubit register we call the {\em auxiliary} or
{\em ancillary register}
is used to induce relative phase changes in the first register.
In view of the second register the function evaluation employs a
translation operator $T_z : \ket{y} \mapsto \ket{y+z}$ where
$z=f(x)$ is dependent of the state of the first register.
That is, $U_f$ can be regarded as an operation
$ \ket{x}\ket{y} \mapsto \ket{x} T_{f(x)} \ket{y} $.
If we concentrate on the ancillary register,
the required operation is
$J_{k,z}: \ket{y} \mapsto \w_M^{kz} \ket{y}$ for all $y\in \Z_M$.
$J_{k,z}$ has an eigenvalue $\w_M^{kz}$ and the corresponding
eigenspace is the whole Hilbert space ${\cal H}_M$.

For simplicity, let us assume that $N$ and $M$ are powers of 2,
that is, $N=2^n$ and $M=2^m$ for some nonnegative integers $n$ and
$m$.
Let $R_{k,I} = \QFT^{-1} T_{-k} \QFT$
where $I$ is an identity map and
$\QFT$ is the quantum Fourier transform.
Then it maps $\ket{y}$ to $\w_M^{ky} \ket{y}$
in which the phase-encoded information depends on the state.

We first describe an algorithm to implement $J_{k,z}$.
We prepare an arbitrary $m$-qubit register with no initialization
and let $\ket{\psi} = \sum_{y=0}^{M-1}a_y\ket{y}$ be its state.
Now we proceed the following steps.

\begin{enumerate}
\item Applying the translation $T_z$ we get
      \[ \sum_{y=0}^{M-1}a_y \ket{y+z}. \]
\item Applying $R_{k,I}$ we obtain
      \[ \sum_{y=0}^{M-1} \w_M^{k(y+z)} a_y \ket{y+z}. \]
\item Applying $T_z^{-1}=T_{-z}$ the state becomes
      \[ \sum_{y=0}^{M-1} \w_M^{k(y+z)} a_y \ket{y}. \]
\item Apply $R_{k,I}^{-1}=\QFT^{-1} T_{k} \QFT$.
      Then the final state is
      \[ \w_M^{k z}\ket{\psi}. \]
\end{enumerate}

This algorithm realizes $J_{k,z}$
via $R_{k,I}^{-1} T_z^{-1} R_{k,I} T_z$.
That is, for an arbitrary initial state $\ket{\psi}$
\[
 R_{k,I}^{-1} T_z^{-1} R_{k,I} T_z
 \ket{\psi} = \w_M^{kz} \ket{\psi}.
\]
The algorithm to implement $J_{k,z}$ is not unique. In fact, all
cyclic rotational permutations of the operational steps are
identical. If we write $[A,B]=ABA^{-1}B^{-1}$, then we can easily
check that $J_{k,z} = [R_{k,I}^{-1},T_z^{-1}] = [T_z,R_{k,I}^{-1}]
 = [R_{k,I},T_z] = [T_z^{-1},R_{k,I}]$ and
$J_{-k,z} = [R_{k,I},T_z^{-1}] = [T_z,R_{k,I}]
 = [R_{k,I}^{-1},T_z] = [T_z^{-1},R_{k,I}^{-1}]$.
For example, we can start at Step 2, perform successive steps, and
end at Step 1. Noting that $S_{k,I} = \QFT T_{-k} \QFT$ maps
$\ket{y}$ to $\w_M^{ky} \ket{-y}$, we can easily check that
\[
 R_{k,I}^{-1} T_z^{-1} R_{k,I} T_z
 = S_{k,I} T_z S_{k,I} T_z .
\]
We remark that $S_{k,I}^{-1} = S_{k,I}$.
Thus we have another algorithm for $J_{k,z}$. However, the number
of $T_z$ or $T_{-z}$ in each implementation is
always equal to or more than two and cannot be reduced.
This is because we require no initialization,
which we shall explain more precisely later.
It follows that the $f$-conditioned phase transform requires two
evaluations of $f$.

Especially when $M=2$ and $k=1$,
$\QFT$ is the Walsh-Hadamard operator $W$ and
$T_z=T_{-z}$ is the Pauli spin matrix
$\sigma_x = \left(
\begin{smallmatrix} 0 & 1 \\ 1 & 0 \end{smallmatrix}
\right)$
which represents a bit-flip operator.
Thus the operator $R_{k,I}$ is just a phase-flip operator
$\sigma_z = \left(
\begin{smallmatrix} 1 & \hphantom{-} 0 \\ 0 & -1 \end{smallmatrix}
\right)$.
The overall scheme \cite{Kim} is $\sigma_z U_f \sigma_z U_f$.

We turn to the operator $R_{k,f}$ for a general function $f$.
In this case we need two registers as we have already mentioned.
In the ancillary register
$U_f$ can be seen as a translation $T_{f(x)}$
conditioned on the control register which state is in $\ket{x}$.
Using the above algorithm
we can perform $R_{k,f}$ without any initialization of
the ancillary register.
We let $\ket{\phi} = \sum_{x=0}^{N-1} a_x \ket{x}$ and
$\ket{\psi} = \sum_{y=0}^{M-1} b_y\ket{y}$
be the states of the control and ancillary registers,
respectively and perform the following algorithm.

\begin{enumerate}
\item Applying $U_f$ we get
      \[ \sum_{x=0}^{N-1} \sum_{y=0}^{M-1}
      a_x b_y \ket{x} \ket{y+f(x)}. \]
\item Applying $I \otimes R_{k,I}$ we obtain
      \[
       \sum_{x=0}^{N-1} \sum_{y=0}^{M-1}
       a_x b_y \w_M^{k(y+f(x))} \ket{x} \ket{y+f(x)}.
      \]
\item Applying $U_f^{-1}=U_{-f}$ the state becomes
      \[
       \sum_{x=0}^{N-1} \sum_{y=0}^{M-1}
       a_x b_y \w_M^{k(y+f(x))} \ket{x} \ket{y}.
      \]
\item Apply $I \otimes R_{k,I}^{-1} = I \otimes R_{-k,I}$.
      Then the final state is
      \[
       \left( \sum_{x=0}^{N-1} \w_M^{k f(x)} a_x \ket{x} \right)
       \ket{\psi} .
      \]
\end{enumerate}

If we discard the ancillary register, then we obtain the
$f$-conditioned phase transform
$R_{k,f} : \ket{x} \mapsto \w_M^{kf(x)} \ket{x}$
without any initialization of ancillary register.
The ancillary register can consist of
any $m$ qubits which may be composed of parts of any other registers
even though they are still being used in another computation
regardless of their states possibly entangled with other qubits.
We note that after extracting the desired relative phase
the initial state of ancillary register is recovered.
Thus this temporary register can be used in continuing
the previously stopped computation.

Our algorithm requires both $U_f$ and $U_{-f}$. In other
words, at least two evaluations of $f$ are necessary.
If we can initialize the ancillary register,
only one evaluation of $f$ is sufficient.
We see that $\QFT \ket{-k}$ is an eigenvector of $T_z$
with the corresponding eigenvalue $\w_M^{kz}$. If we let
$\ket{\psi} = \QFT T_{-k} \ket{0}$, then $U_f : \ket{x} \ket{\psi}
\mapsto \w_M^{k f(x)} \ket{x}\ket{\psi}$.
The special case when $k=1$ was studied in \cite{Cleve1,Cleve2}.

However, if we are to start with any state of ancillary register
we have to find unitary operators $V$ and $W$
satisfying $V T_z W = \w_M^{kz} I$.
Notice that $T_z$ has to be used at least once whether we employ
initialization or not. Since $W^{\dagger} V T_z = \w_M^{kz} I$, it
is enough to find a unitary operator $V$ such that $V T_z =
\w_M^{kz} I$. Since $V = \w_M^{kz} T_{-z}$, $V$ has to depend on
$z$. Thus in some step of the algorithm we have to use information
on $z$ once more and so we need at least two $T_z$ or $T_{-z}$.
Therefore to realize $R_{k,f}$ we need at least two evaluation of
$f$. In this sense the algorithm presented here is optimal.

Let us consider the case $f:\Z_N \rightarrow [0,1) \subset \R$.
Then with $m$-bit approximation
$\tilde{f} : \Z_N \rightarrow \Z_M$ of $f$
the approximate $f$-conditioned phase transform $R_{k,\tilde{f}}$
can be accomplished.
This approximate $f$-conditioned phase transform
is useful in the conditional $\gamma$-phase transform and
the $\beta$-phase diffusion transform which are constructed
in \cite{CK1,CK2}.
Similarly we can achieve any $m$-bit approximate of
more general phase transform which can be described
by $R_{1,g\circ f}$ given a function
$g:\Z_M \rightarrow [0,1) \subset \R$.

\section{Generalized Deutsch-Jozsa Problem}
The Deutsch-Jozsa problem is to determine whether a function $f:\Z_N
\rightarrow \Z_2$ is either constant or balanced
under the assumption that $f$ is either one.
This problem, in which $m=1$ and thus $\w_M=-1$,
can be solved by measuring $W_n R_{1,f} W_n \ket{0^n}$:
when the outcome is $\ket{0^n}$ $f$ is constant and
otherwise $f$ is balanced.
This procedure can easily be extended to solve the generalized
Deutsch-Jozsa problem by employing $f$-conditioned phase
transform.
We say that $f$ is {\em evenly distributed} if $f$ has evenly distributed
$D>0$ values and the numbers of $x$ which map to the same value are all equal.
If $f$ is evenly distributed, then there exist $D>0$ and $a\ge0$ such the
period of the range of $f$ is $L = M/D $ with a possible initial shift $a$;
\[
 \{ f(x) \, :\, x \in \Z_N \}
 = \{ jL +a \, :\, j \in \Z_D \}
\]
and $| A_1 | = |A_2| = \cdots |A_{D}|$
where $ A_j = \{ x \in \Z_N \, :\, f(x) = jL+a \}$ for $j \in \Z_D$.
The {\em generalized Deutsch-Jozsa problem} is to determine
whether $f$ is constant or evenly distributed
when $f$ is either one.

We now explain the procedure to solve the generalized Deutsch-Jozsa
problem.
We prepare an $n$-qubit register with its initial state being $\ket{0^n}$
and apply $W_n R_{k,f} W_n$ for $k\ne 0$
where $W_n$ is the $n$-qubit Walsh-Hadamard transform.
Then we have
\[
\ket{0}
\stackrel{W_n}{\longrightarrow}
\frac{1}{\sqrt{N}}\sum_{x=0}^{N-1} \ket{x}
\stackrel{R_{k,f}}{\longrightarrow}
\frac{1}{N}\sum_{y=0}^{N-1}\w_M^{kf(x)} \ket{x}
\stackrel{W_n}{\longrightarrow}
\left( \sum_{x=0}^{N-1} (-1)^{x \cdot y} \w_M^{kf(x)} \right) \ket{y}
\]
where $x \cdot y$ stands for the XOR of the bitwise AND of the binary strings
$x$ and $y$ in $\Z_2^n$.
Let $S$ be the inner summation in the final state;
\[
 S = \sum_{x=0}^{N-1} (-1)^{x \cdot y} \w_M^{k f(x)} .
\]
If $f$ is constant, then
\begin{eqnarray*}
 S &=& \w_M^{kf(0)} \sum_{x=0}^{N-1} (-1)^{x\cdot y} \\
   &=&
\left\{
\begin{array}{ll}
0 & \mbox{when } y \ne 0 \\ N \w_M^{kf(0)} & \mbox{when } y = 0 .
\end{array}
\right.
\end{eqnarray*}
If $f$ is evenly distributed, then for $y=0$ we have
\begin{eqnarray*}
 S &=& \frac{N}{D}\sum_{j=0}^{D-1} \w_M^{k(jL+a)} \\
   &=& \frac{N}{D}\w_M^a \sum_{j=0}^{D-1} \w_D^{kj} \\
   &=& 0 .
\end{eqnarray*}
Hence when $f$ is constant the final state is $\ket{0}$
while $\ket{0}$ disappears when $f$ is evenly distributed. Now we
measure the register. If the outcome of the measurement is $\ket{0}$
then we conclude that $f$ is constant.
Otherwise, we conclude that $f$ is evenly distributed.
Thus we can solve the generalized Deutsch-Jozsa problem
by a single evaluation of $f$
with known initialization of the ancillary register
and by two evaluations of $f$
with unknown state of the ancillary register.

We note that our procedure is independent of $D$, $L$ and $a$.
To determine whether $f$ is constant or evenly distributed
we need $N/D+1$ evaluations of $f$ classically
in worst case.
This is the case when $D$ or $L$ is known.
However, if neither $D$ nor $L$ is available,
any classical algorithm for this problem would require
$N/2+1$ evaluations of $f$ in the worst case
before determining the answer with certainty.
Whence the generalized Deutsch-Jozsa problem has the same complexity
as the original Deutsch-Jozsa problem.

Furthermore, when $f$ is evenly distributed
we can determine $D$, $L$ and $a$.
The image of $f$ has period $L$ with initial shift $a$.
Finding the period $L$ of a function with an unknown initial shift
$a$ can easily be solved on a quantum computer.
Actually the quantum Fourier transform wipes off the initial shift
and changes its period to $M/L=D$.
This useful property was used to solve
factoring problem by Shor \cite{Shor}.

When $f$ is onto, $f$ is an evenly distributed function
if and only if $f$ is an $r$-to-one function where $r=N/D$.
Thus the generalized Deutsch-Jozsa algorithm can determine
whether $f$ is constant or $r$-to-one.
The $r$-to-one function appears in collision and claw problems \cite{BHT}
under the assumption that $f$ is onto.

We note that for general positive integers $N$ and $M$,
the approximate Fourier transform in \cite{Kitaev} can be used.

\section*{Acknowledgments}
We gratefully acknowledge the support of Research Institute of
Mathematics.

\end{document}